\documentstyle[preprint,aps]{revtex}
\begin{document}
\draft
\preprint{\vbox{To be published in Phys.\ Lett.\ B \hfill SSF94-02-01}}
\title{Two-nucleon knockout contributions to the $^{12}$C$(e,e'p)$
reaction in the dip and \protect{$\Delta$}(1232) regions}
\author{J. Ryckebusch \thanks{Postdoctoral Research Associate NFWO},
V. Van der Sluys, M. Waroquier \thanks{Research Director NFWO}}
\address{Institute for Theoretical Physics,
         Proeftuinstraat 86, B-9000 Gent, Belgium}
\author{L.J.H.M. Kester, W.H.A. Hesselink}
\address{Department of Physics and Astronomy, Free University, De
Boelelaan, 1081 HV Amsterdam, The Netherlands}
\author{E. Jans and A. Zondervan}
\address{Nationaal Instituut voor Kernfysica en Hoge-Energiefysica,
sectie K (NIKHEF-K), \\
         P.O. Box 41882, 1009 DB Amsterdam, The Netherlands}
\date{\today}
\maketitle
\newpage

\begin{abstract}
The contributions from $^{12}$C$(e,e'pn)$ and $^{12}$C$(e,e'pp)$
to the semi-exclusive $^{12}$C$(e,e'p)$ cross section have been
calculated in an unfactorized
model for two-nucleon emission.
We assume direct two-nucleon knockout after
virtual photon coupling with the two-body pion-exchange currents in the
target nucleus. Results are
presented at several  kinematical conditions in the dip
and $\Delta$(1232) regions.   The calculated two-nucleon knockout strength
is observed to
account for a large fraction of the measured $(e,e'p)$ strength
above the two-nucleon emission threshold.
\end{abstract}
\newpage

Rosenbluth separations of the inclusive electron scattering $(e,e')$
cross section have established the predominant transverse nature of the
strength in the dip region between the quasielastic and $\Delta$(1232)
peaks \cite{bar83}.  Whereas the quasielastic and $\Delta$(1232) peaks
have been the subject of an extensive theoretical activity, the physics
of the dip region is less explored and relatively badly understood.
In a Fermi-gas model calculation, Van Orden and Donnelly have shown that
meson-exchange currents (MEC) can make up for a significant
part of measured inclusive
strength in the dip region \cite{ord81}.  Recently, the
Fermi-gas calculations have been readdressed by Dekker, Brussaard and
Tjon in a fully relativistic framework \cite{dek91}.
They find a substantially
larger $(e,e')$ response for energy transfers above the quasielastic
peak than Van Orden and Donnelly, an effect which they attribute to their
non-static treatment of the $\Delta$ current operator.  The predominance
of the MEC was
confirmed in the semi-phenomenological quasi-deuteron calculations for
$^{12}$C$(e,e')$ by Laget and Chr\'{e}tien-Marquet \cite{bar83}.
Although the
quasi-deuteron and the recent Fermi-gas results look promising,
a complete
microscopic description of the inclusive electron scattering data
in the dip region is still lacking.  Given the
important role attributed to MEC, it has been
suggested that two-nucleon knockout may account for a
significant part of the
strength \cite{don78}.
Above the pion-production threshold, also quasi-free pion production
via intermediate $\Delta$(1232) excitation is a
significant source of $(e,e')$ strength \cite{bar83}.

More detailed information about the various reaction mechanisms is
supposed to be gained from coincidence experiments.  The first $(e,e'p)$
measurement in the dip region was reported by Lourie et al.
\cite{lou86}.  The attained $^{12}$C$(e,e'p)$ spectrum is
characterized by peaks corresponding to knockout from the $1p$ and $1s$
shell and by a significant and almost
uniform continuum strength
extending to high missing energies E$_m$=$\omega$-T$_p$. Here, $\omega$
is the transferred energy in the electron-scattering process
and T$_p$ is the measured proton kinetic energy.
The results of the investigations reported in ref.~\cite{lou86} thus
exclude a purely one-body knockout character of the $(e,e'p)$ reaction
process in the dip region.   Similar observations were made
 in the subsequent $^{12}$C$(e,e'p)$ measurements
of Baghaei et al. \cite{bag89} probing the nuclear response in
the $\Delta$-resonance region.
Here, the $E_m$ dependence of the proton strength
is characterized by a strong enhancement above the pion production threshold.

The effect of the final-state interaction (FSI) on the continuum
part of the $(e,e'p)$ spectra was investigated by Takaki
\cite{tak89a,tak89b}. He concluded that FSI effects following virtual
photon coupling with a one-body current does neither qualitatively nor
quantitatively account for the data of ref.~\cite{lou86}.
Based on the results of calculations performed with a simple model for
two- and three-nucleon absorption he argued further that
a reaction mechanism in which the
virtual photon couples with the
two-body currents may dominate the $(e,e'p)$ cross sections for the
missing energy region right above the
two-nucleon emission threshold.  Such a reaction mechanism, however,
was found to generate relatively little strength at
higher missing energies.  In order to account for the observed $(e,e'p)$
strength at high missing energies, Takaki invoked three- and
more-nucleon processes.

Here, we report on a calculation of the two-nucleon
knockout contribution to the semi-exclusive $^{12}$C$(e,e'p)$ reaction
in the dip and $\Delta$(1232) regions and compare the results
with earlier measurements from
MIT-Bates and new data taken at NIKHEF-K.  The latter represent the
first semi-exclusive $(e,e'p)$
data for a complex nucleus taken in non-parallel kinematics.
All MIT-Bates measurements in the dip and $\Delta$ regions are
performed in parallel kinematics, i.e. the proton is detected along
the direction of the transferred momentum {\bf q}.

The two-nucleon knockout contribution to the semi-exclusive $(e,e'p)$
channel is determined by integrating the $(e,e'pp)$ and $(e,e'pn)$ cross
sections over the solid angle of the undetected hadron :
\begin{eqnarray}
{d^6 \sigma \over dE_p d \Omega _p d\epsilon ' d \Omega _{\epsilon '}}
(e,e'p) & & =   \sum_{f} \int d \Omega_{p'}
 {d^8 \sigma \over dE_p d \Omega _p  d \Omega _{p'}
d\epsilon ' d \Omega _{\epsilon '}}
(e,e'pp) \nonumber \\ & & +
\sum_{f} \int d \Omega_{n}
 {d^8 \sigma \over dE_p d \Omega _p  d \Omega _{n}
d\epsilon ' d \Omega _{\epsilon '}}
(e,e'pn)\;,
\label{coneep}
\end{eqnarray}
where the sum over $f$ extends over all final states of the
residual (A-2) system.
Here, it is assumed
that the two-hadron knockout process is the result of a direct
knockout (DKO) mechanism after virtual photon absorption on a correlated
nucleon pair in the target nucleus (fig.~1).  This means that the
photoabsorption mechanism involves two active nucleons.
The term ``correlations'' is defined here in its
most general sense, referring to any type of mutual interaction two
nucleons experience when embedded in a nuclear medium. The calculations
presented here are restricted to rather moderate momentum transfers q
and therefore one
may assume that the nucleon-nucleon correlations are dominated by one-pion
exchange mechanisms.  As illustrated in fig.~1 we consider all diagrams
with one exchanged pion retaining both the non-resonant (fig.~1(a) and
1(b)) and resonant (fig.~1(c)) terms.  Accordingly, in the calculation
of the $(e,e'NN)$ cross sections the virtual photon is taken to
couple exclusively with a transverse two-body current.  This means that
we discard all longitudinal contributions to the $(e,e'NN)$ cross
sections.  The longitudinal strength can be expected to be dominated by the
short-range correlations \cite{carlotta} and multi-step processes
\cite{tak89b}.  Given that the $(e,e')$ strength has a
predominantly transverse nature
in the dip and $\Delta$ regions,
one may infer that the two-body currents of fig.~1 are responsible for
a large fraction of the $(e,e'p)$ strength.


Following standard
procedures the coincidence differential cross section for
the \mbox{A$(e,e'NN)$A-2} process
in the laboratory frame can be written as :
\begin{equation}
{d^8 \sigma \over dE_b d \Omega _b d \Omega _a d \epsilon ' d \Omega
_{\epsilon '}} (e,e'N_aN_b) =
{1 \over 4 (2\pi)^8 } k_a k_b E_a E_b f_{rec} \sigma_{M}
\left[ v_T W_T + v_S W_{TT} \right] \;,
\label{eepnn}
\end{equation}
where  $\sigma _M$ is the Mott cross section
and $f_{rec}$ a recoil
factor.  The electron kinematical factors are given by
$v_T=tg^2{\theta _e \over 2} -  {q_{\mu}^2 \over 2 {\bf q}^2}$ and
$v_S={q_{\mu}^2 \over 2 {\bf q}^2}$.
Notation conventions for the kinematical variables are summarized
in fig.~1. The structure functions $W$ are defined
in terms of the transition matrix elements $m_{fi}$ of the transverse
two-body current operator :
\begin{eqnarray}
W_T & = & \left|m_{fi}(\lambda =+1) \right|^2 + \left|m_{fi}(\lambda =-1)
 \right|^2 \nonumber \\
W_{TT} & = & 2 Re \left[ \left(m_{fi}(\lambda =+1) \right) \left(m_{fi}(\lambda
=-1)
\right)^* \right] \;,
\end{eqnarray}
and constitute the essential quantities to be calculated.  Remark that
eq.~(\ref{eepnn}) has been derived under the assumption that electro-induced
two-nucleon ejection is a purely transverse process.
For the reactions under consideration where two hadrons
characterized by the momentum/spin variables $({\bf k}_a,m_{s_{a}})$ and
$({\bf k}_b,m_{s_{b}})$ are ejected, the transition matrix element reads :
\begin{equation}
m_{fi}(\lambda)=\left< {\bf k}_a {1 \over 2} m_{s_{a}} ;
{\bf k}_b {1 \over 2} m_{s_{b}} ; J_R M_R (A-2)
\left| J_{\lambda}^{[2]}(q) \right| g.s.(A) \right> \qquad \qquad
(\lambda=\pm 1)\;,
\label{feynman}
\end{equation}
with $J_{\lambda}^{[2]}$ the Fourier transformed nuclear two-body current.
In the evaluation of the diagrams 1(a) and 1(b) we
employ the non-relativistic
reduction of the current operators that correspond with the one-pion
exchange potential.  They are derived from the pseudovector $\pi NN$
coupling Lagrangian \cite{ord81,riska}.
The
non-static $\Delta$ current operator we use in our investigations is :
\begin{eqnarray}
{\bf J}^{(\pi \bigtriangleup)}({\bf q};{\bf k}_1,{\bf k}_2) & = &
\frac {2 i f_{\gamma N \bigtriangleup} f_{\pi N \bigtriangleup} f_{\pi
NN}} {9 m_{\pi}^3 (M_{\Delta} - M_N - \omega -{i \over 2}
\Gamma_{\Delta}(\omega) )}
\left\{ \left[ -\left(\mbox{\boldmath$\tau$}_1 {\bf \times}
\mbox{\boldmath$\tau$}_2
\right)_z {\mbox{\boldmath$\sigma$}_2 {\bf \cdot} {\bf k}_2 \over
k_2^2+m_{\pi}^2}
(\mbox{\boldmath$\sigma$}_1 \times
{\bf k}_2) \times {\bf q} \  +
 \right. \right. \nonumber \\ +
& & \left. \left . 4 (\mbox{\boldmath$\tau$}_2)_z
{\mbox{\boldmath$\sigma$}_2
{\bf \cdot} {\bf k}_2 \over k_2^2+m_{\pi}^2}
({\bf k}_2 {\bf \times} {\bf q})
 \right]
+ 1 \longleftrightarrow 2 \right\} \;,
\label{eq:pidelta}
\end{eqnarray}
which corresponds with the non-relativistic reduction of the $\Delta$
current operator as derived in ref.~\cite{dek91}.  The $\Delta$-width
$\Gamma _ {\Delta} (\omega)$ was taken according to
the parametrization of Oset,
Toki and Weise \cite{ose82}.  The various coupling constants are ${f_{\pi
NN} ^2 \over 4 \pi}$=0.079, ${f_{\pi N \Delta} ^2 \over 4 \pi}$=0.37 and
$f_{\gamma N \Delta} ^2 $=0.014.   For the electromagnetic formfactors
the standard dipole form is adopted.  For the strong $\pi NN$ and $\pi N
\Delta$ formfactor a monopole form $(\Lambda _ \pi^2-m_ \pi^2)/
(\Lambda _\pi^2+{\bf k}^2)$
with a cut-off mass $\Lambda _\pi$ fixed at  1200~MeV was used.

In order to account for the distortions that the escaping nucleons
undergo through the interaction with the (A-2) spectator nucleons
a technique based on a partial wave expansion for the wave
function of each of the escaping particles has been developed.
This procedure is a natural
extension of the shell-model approach to one-nucleon knockout reactions
\cite{mahaux}.  Essentially, the final antisymmetrized state with two
escaping nucleons having momenta ${\bf k}_a$ and ${\bf k}_b$ is obtained
by performing an expansion in terms of two-particle two-hole ($2p-2h$)
eigenfunctions of a mean-field potential.  More details can be found
elsewhere \cite{ryc94}.  Since the final state is expressed
in coordinate space, the integrations over the solid angle of the
undetected hadron in eq.~(\ref{coneep}) can be performed
analytically, thus keeping
the numerical calculations feasible.  A
similar procedure as the one sketched here for the semi-exclusive
$(e,e'p)$ reaction has been worked out in more detail for the semi-exclusive
$(\gamma,p)$ reaction in ref.~\cite{ryc94a}.  In the context of
$(\gamma,NN)$ reactions, the two-nucleon emission
model adopted here has been shown to give
 a fair account of the absolute $^{16}$O$(\gamma,pn)$ and
$^{16}$O$(\gamma,pp)$ angular cross section data \cite{ryc94a}.

In the $^{12}$C$(e,e'p)$ calculations all possible contributions from the
removal of proton-neutron and proton-proton pairs in the (1p$_{3/2}$)$^2$,
(1p$_{3/2}$,1s$_{1/2}$) and (1s$_{1/2}$)$^2$
shell-model configurations are included.  The
single-particle wave functions and scattering phase shifts entering the
matrix elements are obtained from a Hartree-Fock calculation with an
effective Skyrme type of interaction. Within the adopted model
assumptions,  the residual
nucleus is created in a $2h$ configuration $\mid (hh')^{-2}>$
relative to the ground state of
the target nucleus. Therefore, all spectroscopic
information in the calculations is contained in the two-hole spectral
function $S_{hh'}$.  The $S_{hh'}(E_x)$ determines the distribution of the
$\mid (hh')^{-2}>$ strength as a function of the
excitation energy E$_x$ in the (A-2)
nucleus. This distribution was parametrized according to \cite{ryc94a} :
\begin{equation}
S_{hh'}(E) = \int _{0}^{E} S_{h}(E')S_{h'}(E-E')dE' \;.
\end{equation}
The single-hole spectral functions $S_h(E)$ occurring in this expression
are taken from ref.~\cite{jeukenne}.

In fig.~\ref{angdepen} we present the
calculated  $^{12}$C$(e,e'pp)$ and $^{12}$C$(e,e'pn)$ contributions to
the semi-exclusive $^{12}$C$(e,e'p)$ channel as a function of the
missing energy at different values of the proton angle $\theta _p$.
(-180$^{\circ} \leq \theta_p \leq$ 180$^{\circ}$). The
proton scattering angle $\theta _p$ is expressed relative to the direction of
${\bf q}$.   Since
the $(e,e'NN)$ reaction is a purely transverse process
within the adopted model assumptions, the azimuthal dependence of
the calculated semi-exclusive cross sections is solely
determined by $cos(2\varphi _p)$.  All forthcoming considerations are for
in-plane kinematics ($\varphi _p = 0^{\circ}, 180^{\circ})$.  Accordingly,
the angular dependence of the calculated cross
sections is uniquely determined by
the absolute proton angle $\mid \theta _p \mid$.
The electron kinematics of fig.~\ref{angdepen}
are taken from a recent experiment
peformed at
NIKHEF-K and are typical for the dip region \cite{leo94}.
For these kinematics, the numerical calculations predict the
two-hadron emission strength to
sharply rise above the threshold and to be spread over a wide range of
missing energies.  The calculated missing energy spectra show
a clear dependence on $\theta _p$.  The $E_m$
strength distribution, which has a clear bump structure in parallel
kinematics ($\theta _p=0^{\circ}$), exhibits a wider structure
with increasing
proton detection angle $\mid \theta _p \mid$.
In conformity
with the conclusions drawn by Takaki, our calculations suggest that in
parallel kinematics a uniform $(e,e'p)$ strength distribution extending
over a wide $E_m$ range is incompatible with two-nucleon knockout as the
sole contributing channel.

Inspecting fig.~\ref{angdepen} it is clear that
the $pp$ and $pn$ contributions  show a slightly different functional
dependence on $\theta _p$.  The $pp$ part decreases more rapidly with
increasing $\theta _p$ than the $pn$ part.  This property is reflected
in the $(e,e'pn)$ to $(e,e'pp)$ ratio which is less than ten in parallel
kinematics and steadily grows as the proton is detected at larger
angles $\mid \theta _p \mid$.


The
calculated cross sections
have been compared with the results of $^{12}$C$(e,e'p)$
experiments recently performed at NIKHEF-K and earlier measurements by
Baghaei {\em et al.} \cite{bag89}.  The measurements at NIKHEF were
performed at two sets of
values for the energy and momentum transfer, one in the
dip region ($\omega$=212~MeV, q=270~MeV/c) and the other one at the rising
slope of the $\Delta$-resonance peak ($\omega$=263~MeV, q=303~MeV/c).
Here, only a selection of the NIKHEF data is shown.
An in-depth comparison of the predictions shown in
fig.~\ref{angdepen} and the NIKHEF
data at $\omega$=212~MeV, which include missing-energy spectra at
various proton angles, will be presented elsewhere \cite{leo94}.
Summarizing, for the kinematical conditions of fig.~\ref{angdepen} the
calculations essentially reproduce the data for $\mid \theta _p \mid
\geq 74^\circ$ and underestimate the data by about a factor of two at
smaller values of $\mid \theta _p \mid$.
In figs.~\ref{leon2}  and \ref{baghaei} the calculated $E_m$ spectra
are compared with the data which are representative for the
low-energy side of the $\Delta$(1232) resonance peak.
For both data sets the transferred energy $\omega \approx $ 270~MeV.
The MIT-Bates
results of fig.~\ref{baghaei}
were taken at slightly larger momentum transfer (q=401~MeV/c) than the
NIKHEF-K data of fig.~\ref{leon2} (q=303~MeV/c).
The combined data sets
cover a range of
proton angles from $\mid \theta _p \mid =0^{\circ}$ up to $\mid \theta
_p \mid =113^{\circ}$.
Generally, a
reasonable description of the low $E_m$ part of the  $^{12}$C$(e,e'p)$
spectra is achieved.  For $\theta _p=0^{\circ}$ and $38^{\circ}$ the
theory clearly falls short of accounting for the measured strength at
high $E_m$.  This is expected since the model does not include the
real-pion production channels.  As the mismatch
between ${\bf q}$ and ${\bf k}_p$ is large,
pion electroproduction $(e,e'p \pi^-)$ is not expected
to contribute substantially to the proton
spectra at large $\theta _p'$s.  As such, the backward proton angles are
best suited to study the two-nucleon knockout contribution to the
semi-exclusive processes.   It is precisely for
$\theta _p=113^{\circ}$ that we arrive at a fair description of both the
$E_m$ dependence and the magnitude of the data.  Therefore, it is
tempting to conclude that at the backward angles
the semi-exclusive cross sections are dominated by
two-nucleon knockout caused by MEC and $\Delta$
currents.
Remark further that the
calculated missing-energy
spectra of figs. \ref{leon2} and \ref{baghaei} show a functional
dependence on $\theta _p$ similar to the curves of fig.~\ref{angdepen} which
are obtained at a lower $\omega$.

Summarizing, the $(e,e'pp)$ and $(e,e'pn)$
contribution to the semi-exclusive
$^{12}$C$(e,e'p)$ process in the dip and $\Delta$
regions has been computed using a
mean-field approximation to account for the distortions that the ejected
hadrons undergo.  The electro-induced two-hadron emission processes are
assumed to be caused by a direct knockout
 mechanism following electron scattering
off the two-body pion-exchange and $\Delta$
currents in the target nucleus. The overall shape
of the calculated $^{12}$C$(e,e'p)$ missing energy spectra is found to be
dependent on the proton emission angle $\theta _p$
and as such measurements of the
$(e,e'p)$ specta at several values of $\theta _p$ are well suited
to provide a
deeper insight into the relative importance of the two-nucleon knockout
contributions.  The presented model, which is parameter-free,
 works reasonably well in reproducing that part of the
$^{12}$C$(e,e'p)$ missing energy spectra which resides below the pion
production threshold.    Consequently, two-hadron knockout is found to
be a
substantial contribution to the  $(e,e')$ reaction mechanism
above the
quasielastic peak.   We believe that semi-exclusive $(e,e'p)$
reactions constitute an important tool in the study of
electro-induced two-hadron knockout processes and
provide a direct way of gaining deeper insight into the nature of
nucleon-nucleon correlations in finite nuclei.

{\bf Acknowledgement}

This work was supported by the Belgian
National Fund of Scientific Research (NFWO), the
Stichting Fundamenteel Onderzoek der Materie (FOM) and
the Nederlandse Organisatie voor Wetenschappelijk
Onderzoek (NWO). The authors are grateful
to Dr.~L.B.~Weinstein for kindly providing us with the data files of the
MIT measurements.


\newpage

\begin{figure}
\caption{Schematic representation of an electro-induced two-nucleon
ejection process in a direct knockout picture.  Also the different types
of nucleon-nucleon correlations retained in the calculations are shown :
(a) Seagull, (b) pion-in-flight and (c) $\Delta$-resonance diagrams.}
\label{scheme}
\end{figure}

\begin{figure}
\caption{Theoretical missing-energy spectrum of the $^{12}$C$(e,e'p)$
process at $\epsilon$=475~MeV, $\omega$=212~MeV and q=270~MeV/c.  The
calculated contribution from $(e,e'pn)$ (top) and $(e,e'pp)$ (bottom) is
shown for $\theta _p$=0$^{\circ}$ (solid curve), $\theta
_p$=60$^{\circ}$  (dotted curve), $\theta _p$=120$^{\circ}$ (dot-dashed curve)
and $\theta _p$=180$^{\circ}$ (dashed curve).}
\label{angdepen}
\end{figure}

\begin{figure}
\caption{ Missing-energy spectrum of the $^{12}$C$(e,e'p)$ process
at $\epsilon$=478~MeV, $\omega$=263~MeV and q=303~MeV/c.  The dashed (dotted)
curve shows the calculated $(e,e'pn)$ ($(e,e'pp)$) contribution, the solid
curve their incoherent sum.  The
arrow indicates the threshold for the (e,e$'$p$\pi^-$) channel.}
\label{leon2}
\end{figure}

\begin{figure}
\caption{As in fig.~\protect{\ref{leon2}} but at $\epsilon$=460~MeV,
$\omega$=275~MeV and
q=401~MeV/c. The data are from ref. \protect{\cite{bag89}}.}
\label{baghaei}
\end{figure}

\end{document}